\documentclass[twocolumn,english,aps,showkeys,showpacs]{revtex4-1}
\usepackage[T1]{fontenc}
\usepackage[latin9]{inputenc}
\usepackage{float}
\usepackage{textcomp}
\usepackage{amstext}
\usepackage{graphicx}
\usepackage{babel}

\makeatletter

\pdfpageheight\paperheight
\pdfpagewidth\paperwidth

\newcommand{\lyxmathsym}[1]{\ifmmode\begingroup\def\b@ld{bold}
  \text{\ifx\math@version\b@ld\bfseries\fi#1}\endgroup\else#1\fi}

\@ifundefined{textcolor}{}
{%
 \definecolor{BLACK}{gray}{0}
 \definecolor{WHITE}{gray}{1}
 \definecolor{RED}{rgb}{1,0,0}
 \definecolor{GREEN}{rgb}{0,1,0}
 \definecolor{BLUE}{rgb}{0,0,1}
 \definecolor{CYAN}{cmyk}{1,0,0,0}
 \definecolor{MAGENTA}{cmyk}{0,1,0,0}
 \definecolor{YELLOW}{cmyk}{0,0,1,0}
 }

\makeatother

\begin{document}

\title{Dipolar interactions in magnetic nanowires aggregates.}

\author{Thomas Maurer, }
\affiliation{Laboratoire de Nanotechnologie et
d\textquoteright{}Instrumentation Optique, ICD CNRS UMR STMR 6279,
Universit\'{e} de Technologie de Troyes, BP 2060, 10010 Troyes
cedex, France} \email{thomas.maurer@utt.fr}

\author{Fatih Zighem, Weiqing Fang, Fr\'{e}d\'{e}ric Ott, Gr\'{e}gory Chaboussant}
\affiliation{Laboratoire Léon Brillouin, IRAMIS, CEA-CNRS UMR 12,
CE-Saclay, 91191 Gif sur Yvette, France}

\author{Yaghoub Soumare, Kahina Ait Atmane and Jean-Yves Piquemal}
\affiliation{ITODYS Universit\'{e} Paris 7 - Denis Diderot, UMR CNRS
7086, 2, place Jussieu F-75251 Cedex 05 Paris, France}

\author{Guillaume Viau}
\affiliation{Universit\'{e} de Toulouse, LPCNO, INSA, UMR CNRS 5215,
135, avenue de Rangueil, F-31077 Toulouse Cedex 4.}

\keywords{magnetic nanowires, aggregates, micromagnetic simulations, magnetization
curves, dipolar interactions}

\pacs{75.50.Ww, 75.50.-y, 75.75.-c, 81.07.Gf}

\begin{abstract}
We investigate the role of dipolar interactions on the magnetic properties
of nanowires aggregates. Micromagnetic simulations show that dipolar
interactions between wires are not detrimental to the high coercivity
properties of magnetic nanowires composites even in very dense aggregates.
This is confirmed by experimental magnetization measurements and Henkel
plots which show that the dipolar interactions are small. Indeed,
we show that misalignment of the nanowires in aggregates leads to
a coercivity reduction of only 30\%. Direct dipolar interactions between
nanowires, even as close as $2\,\textrm{nm}$, have small effects
(maximum coercivity reduction of $\sim$15\%) and are very
sensitive to the detailed geometrical arrangement of wires. These
results strenghten the potential of magnetic composite materials based
on elongated single domain particles for the fabrication of permanent
magnetic materials.
\end{abstract}
\maketitle

\section{introduction}

Taking advantage of the shape anisotropy for the fabrication of high
coercivity materials has been considered in the past. We can mention
the development of elongated single domain (ESD) particles
\cite{mendelsohn1955,falk1966,craik1967} or the AlNiCo materials
\cite{AlNiCo} in which elongated magnetic structures are formed by a
proper metallurgical process. These materials are however limited in
their performances to coercive fields of the order of a few mT.
During the last decade it has been shown that elongated magnetic
nano-objects with well defined shapes could be synthesized
\cite{Ung2005,Ung2007,Soumare2008,Soumare2009_Co,Soulantica09}. We
have considered the use of magnetic nanorods for the fabrication of
high magnetic energy polymer composite materials. We have shown that
rather high coercive fields, up to $600\,\textrm{kA/m}$ (=
$0.75\,\textrm{T}$) could be obtained \cite{Maurer_APL}.  However,
the nanowires are not optimally dispersed in the matrix material and
tend to form aggregates (Figure 1). The scope of this communication
is to address the role of the dipolar interactions in nanowire
aggregates on the coercive field in our composite materials.

The role of dipolar interactions in assemblies of magnetic
nanoparticles has already been addressed in several situations. In
some geometries, such as planar arrays of magnetic nanowires
fabricated in porous alumina membranes, the dipolar fields play a
very important role. It has been experimentally demonstrated that
the demagnetizing field $H_{d}$ is simply proportionnal to the
packing density $P$: $H_{d}=-P.M_{s}$
\cite{EncinasPRB01,Zighem2011}. Since $P$ can reach values up to
30\%, the demagnetizing fields are very large and significantly
reduce the coercivity and more generally the anisotropy properties
of the arrays \cite{NielschAPL,Zighem2011}.

\begin{figure}[h]
\includegraphics[bb=70bp 550bp 500bp 780bp,clip,width=8cm]{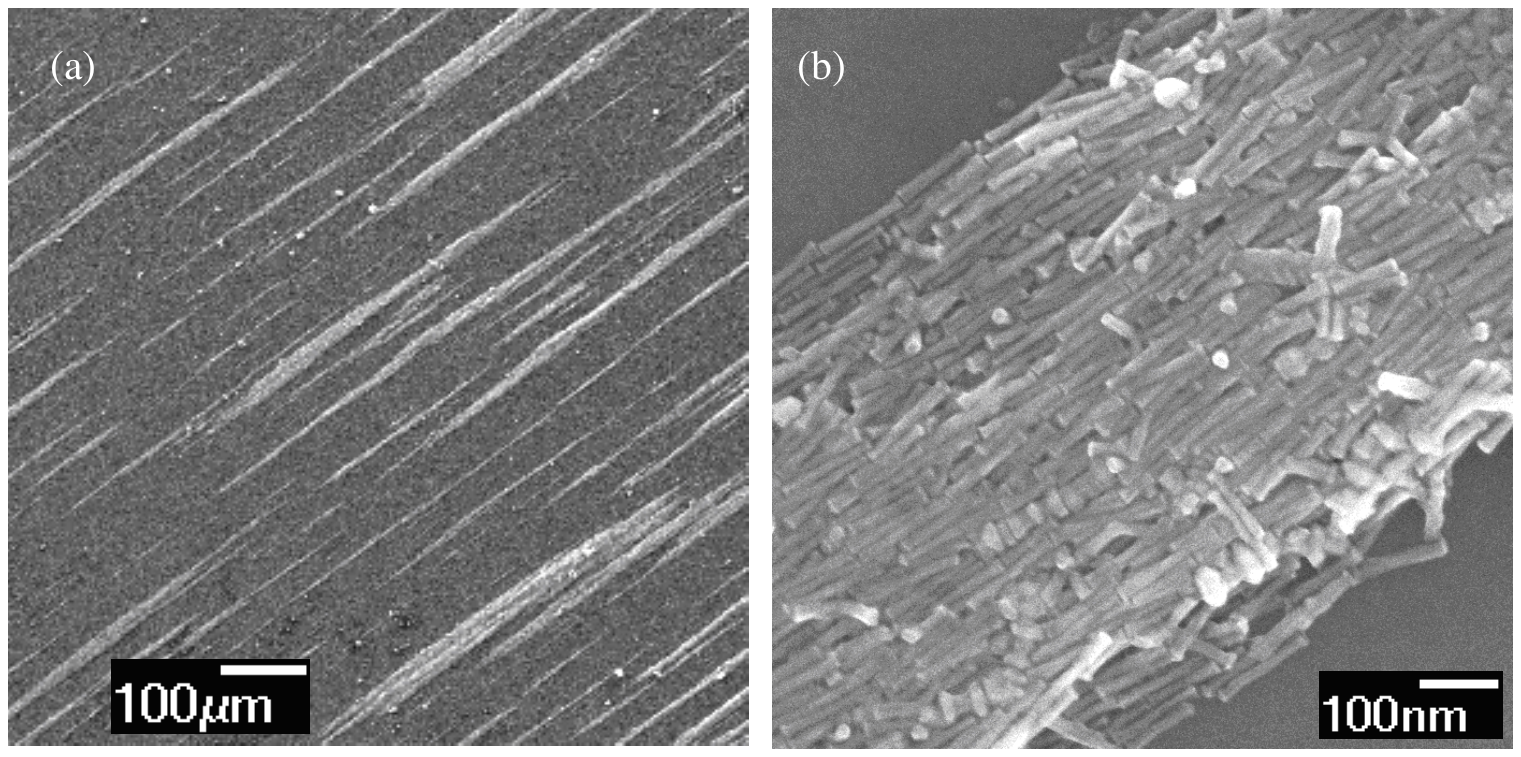}
\label{mesh} \caption{Scanning Electron Microscope images of $Co$
nanowires deposited under a magnetic field on a $Si$ substrate. (a)
At the microscopic scale, the objects form microstructures with
\textit{\emph{cigar}}\emph{ }shapes. (b) Inside these microscopic
structures, the $Co$ nanowires are rather well aligned with an
angular dispersion of the order of $\sigma=7\text{\textdegree}$. }
\end{figure}

The effects of dipolar fields have also been considered for granular
hard magnetic materials \cite{Dobrynin_APL09}. In this case, Henkel
plots \cite{Henkel} -obtained from isothermal remanent magnetization
and dc demagnetization experiments- allow to put into evidence
interactions between magnetic grains \cite{Garci_Henkelplot}. In the
above situations macroscopic demagnetizing field effects must be
taken into account. In our composite materials, the situation is
qualitatively different since elongated cigar shape aggegrates are
formed. Thus macroscopic demagnetizing field effects, within a
cigar, are expected to be negligible. In this communication we focus
on local dipolar field effects between individual wires and we use
micromagnetic simulations to quantify the role of the magnetic
interactions between nanowires on the coercive field.

\section{Magnetic nanowires and aggregates }

We consider the experimental case of $Co$ and $Co_{80}Ni_{20}$ nanowires
synthesized via the polyol process \cite{Ung2005,Ung2007,Soumare2008,Soumare2009_Co}.
These nanowires exhibit diameters in the range $7-15\,\textrm{nm}$
and lengths in the range $100-250\,\textrm{nm}$. In the previous
studies, these nanowires, either randomly oriented or well-aligned
in solvents, have been magnetically characterized. The experimental
hysteresis cycles have been discussed within the Stoner-Wohlfarth
model \cite{Maurer_APL}. The reversal mechanism is usually not a
coherent one and it is initiated at the wires tips. Micromagnetic
simulations put into evidence the role of the detailed shape of these
objects in the magnetization reversal \cite{Ott_JAP2009}. The chemical
synthesis of magnetic wires has been optimized to achieve a cylindrical
shape so as to optimize the shape anisotropy \cite{Soumare2008}.

When the nanowires are dispersed in solvents and dried under a magnetic
field, they tend to form microscopic aggregates (Fig. 1)
so that there can be direct magnetic dipolar interactions between
the objects. The nanowires are covered with an oxide shell which is
typically $1-2\,\mathrm{nm}$ thick \cite{Maurer_PRB} so that the
minimal distance between the ferromagnetic cores is of the order of
$2-4\,\mathrm{nm}$.

Figure 2 represents the dipolar stray field at the tip of
a nanowire ($L=100\,\textrm{nm}$, $r=5\,\textrm{nm}$, $M_{s}=1\,\textrm{T}$)
calculated using FEMM \cite{FEMM}. It illustrates that the dipolar
field is very localized around the tip (in a volume with a typical
size given by the radius of the nanowire). The stray fields are of
the order of $0.1\,\textrm{T}$ at a distance of 4-7nm from the nanowire
tip (see Fig. 2b). Thus, since the nanowires are separated
by distances which can be as small as $2-4\,\textrm{nm}$, the dipolar
field radiated by one wire on its neighbors can still be of the order
of a fraction of one Tesla.

\begin{figure}[h]
\includegraphics[bb=150bp 530bp 460bp 730bp,clip,width=8cm]{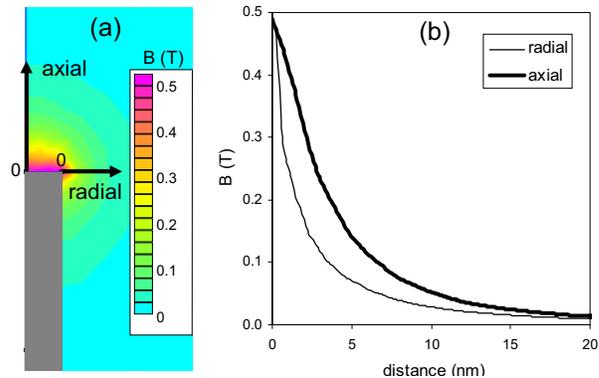}
\label{mesh} \caption{Dipolar field outside the tip of a
($L=100\,\textrm{nm}$, $r=5\,\textrm{nm}$, $M_{s}=1\,\textrm{T}$)
magnetic nanowire (gray area) calculated using FEMM \cite{FEMM}. (a)
Axial mapping of the induction outside the wire. (b) Magnitude of
the induction along the wire axis and the radial axis. The dipolar
stray fields drop quickly but are still of the order of
$0.1\,\textrm{T}$ at a distance of $4-7\,\textrm{nm}$ of the wire
edges.}
\end{figure}

In the present paper, we address the question whether global or local
stray fields may significantly affect the magnetic properties of nanowire
aggregates. Several effects may play a role in modifying the overall
coercivity of nanowires assemblies. They can be classified in the
following way:

(i) The wires may be disoriented with respect to each other. This
leads to a significant loss in the macroscopic coercivity.

(ii) The fact that the wires are assembled in cigar shaped structures
has in principle the effect of decreasing the macroscopic demagnetizing
field effects and increasing the overall coercivity.

(iii) The local stray fields created by a wire tip may create a local
nucleation point and promote the reversal of a neighboring wire.

(iv) The presence of \textquotedblleft{}defective wires\textquotedblright{}
with low coercivity because of a poorly defined shape, a strong misorientation,
or a structural defect may play the role of a nanoscopic nucleation
point inside the cigar microstructure.

In the following, we quantify these different effects in order to
estimate the requirements for obtaining composite materials with optimum
coercivity properties.

\section{Modelling }

The model objects which are considered are cylindrical objects of
length $L=100\,\textrm{nm}$ and radius $r=5\,\textrm{nm}$. The magnetic
parameters used correspond to $hcp$ cobalt epitaxial thin films \cite{tannenwald1961},
saturation magnetization $M_{S}=1400\:\mathrm{kA.m^{-1}}$, exchange
constant $A=1.2\times10^{-11}\mathrm{\, J/m}$. The micromagnetic
calculations have been performed using the \emph{Nmag} package \cite{fishbacher2007}.
The objects were meshed using \emph{Netgen} \cite{NetGen} with meshes
consisting of about 1000 points so that the distance between two nodes
is of the order of the exchange length $\ell_{ex}=\sqrt{2A/\mu_{0}M_{S}^{2}}\approx3.2\:\mathrm{nm}$.
Increasing the finesse of the meshes did not change the results. Magnetocrystalline
anisotropy was not included in the calculation on purpose so as to
avoid masking the dipolar effects by extra anisotropy effects. Adding
an extra magnetic anisotropy to the wires would not change the conclusions.

\subsection{Orientation effects }

For perfectly aligned infinite ellipsoids, the expected coercive field
(obtained from the shape anisotropy) is $\mathrm{M_{S}/2}$ that is
$700\,\mathrm{kA/m}$ ($\equiv0.88T$) in the considered experimental
situation. However, in the case of nanowires of finite length instead
of infinite ellipsoids, a significant drop in the shape anisotropy
is expected ($\sim$25\%)\cite{Ott_JAP2009}. On top of that, the
misorientation of the nanowires induces a loss in the effective coercivity.
We assume a Gaussian distribution. Let $\sigma$ be the width of the
distribution of the wires orientations with respect to the applied
field. $H_{c}$ drops from $0.38\,\mathrm{M_{S}}$ for a very narrow
orientation distribution ($\sigma=1\text{\textdegree}$) down to $0.21\,\mathrm{M_{S}}$
for a random orientation distribution (see Fig. 3). In the observed
situation of Figure 1, the angular distribution of wires has a width
of about $\sigma=7\lyxmathsym{\textdegree}$ leading to a moderate
loss of coercivity ($\sim15\%$) compared to $\ensuremath{\sigma=1\text{\textdegree}}$.
Thus extremely well aligned nanowires do not significantly boost the
performances of the material.

\begin{figure}[h]
\includegraphics[bb=100bp 430bp 520bp 720bp,clip,width=8cm]{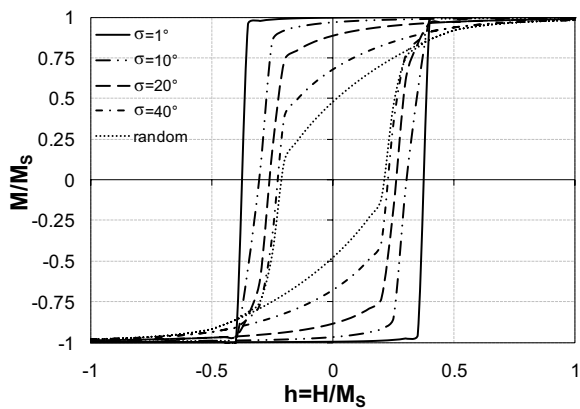}
\label{mesh} \caption{Hysteresis cycles for different wires
orientation distributions: $\ensuremath{\sigma=1\text{\textdegree,
10\textdegree, 20\textdegree, 40\textdegree},}\infty$(random). In
the experimental situation of Figure 1, $\sigma$ has been estimated
at 7\textdegree{}.}
\end{figure}

\subsection{Demagnetizing field effects }

The assembly of the wires into cigar shape structures potentially
modifies the overall demagnetization field distribution. To account
for this effect we have compared the following 3 situations: (A1)
an isolated wire, (A2) a chain of 2 wires separated by $l=2\,\mathrm{nm}$,
(A3) a chain of 3 wires (Fig. 4). The separation of 2 nm which is
a minimum value, corresponds to the experimental situation because the
wires are surrounded by a shell of non magnetic material which prevents
a direct contact between the wires. The reference value of the coercive
field for an isolated wire is $H_{C}\approx375\,\textrm{kA/m}$. The
micromagnetic simulations show that no measurable change in the coercive
field is observed when the wires are arranged in chains. This could
be expected since the wires already have a large aspect ratio so that
we are very close to the case of infinite cylinders. The demagnetizing
coefficient $N$ of a ellipsoid of aspect ratio 10 is equal to 0.02
which is very close to the value $N=0$ for an infinite cylinder.
Increasing further the overall aspect ratio does not induce measurable
effects. When the magnetic field is applied at an angle (ranging from
0\textdegree{} to 20\textdegree{}) with respect to the chains, no
measurable change in the coercivity are observed between the A1, A2
and A3 configurations.

\begin{figure}[h]
\includegraphics[bb=120bp 620bp 450bp 720bp,clip,width=8cm]{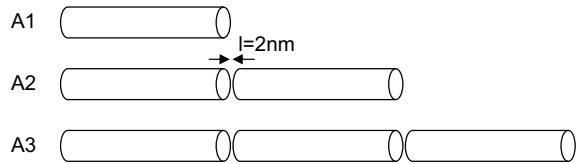}
\label{mesh} \caption{Wires configurations used in the calculations.
(A1) A single wire which is the reference situation. (A2) A chain of
2 wires separated by 2 nm. (A3) A chain of 3 wires. }
\end{figure}

\subsection{Effect of local stray fields }

We have considered several situations in which the stray field of
a wire tip may alter the reversal of a neighboring wire. We have
considered 2 main situations: (i) the case where the stray field acts
at a wire tip (B1 - B2), (ii) the case were the stray field acts on
a neighboring wire edge (B3). The different configurations are illustrated
on Figure 5. We first focused on the situation of aligned nanowires
before considering misalignment which is recurrent in aggregates.
The angle between the two nanowires long axes is noted $\alpha$.

\begin{figure}[h]
\includegraphics[bb=100bp 450bp 350bp 730bp,clip,width=8cm]{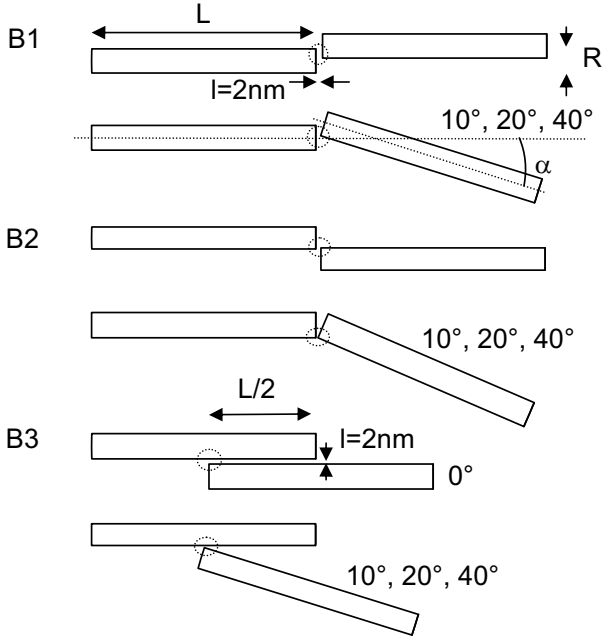}
\label{mesh} \caption{Wires configurations used in the calculations.
(B1) Two wires touching at their tips but shifted by R. The relative
orientation of the wires can vary from $\alpha=0\text{\textdegree}$
to 40\textdegree{}. (B2) Two wires touching at their tips but
shifted by 2R. (B3) A wire creating a stray field around the center
of another wire. }
\end{figure}

Figures 6(a), (b) and (c) respectively display the hysteresis cycles
of the $B1$, $B2$ and $B3$ configurations for different values
of the $\alpha$ angle. For the $B1$ and $B2$ configurations with
$\alpha=0$, the hysteresis cycles are identical to the one of a single
nanowire and $H_{C}\approx375\,\textrm{kA/m}$ (Fig. 6a and b). It
implies that for these configurations -close to $A2$- there is no
measurable effect of the local stray field on the wire tips. Both
wires switch at the same time. The situation differs for the $B3$
configuration where the stray field acts on a neighboring wire edge.
In this case, the coercivity is lowered to $H_{C}\approx325\,\textrm{kA/m}$,
that is a loss of 13\%, while the cycle remains almost perfectly square
(Fig. 6c). The magnetization reversal of both nanowires is simultaneous.

\begin{figure}[h]
\includegraphics[bb=100bp 200bp 360bp 730bp,clip,width=8cm]{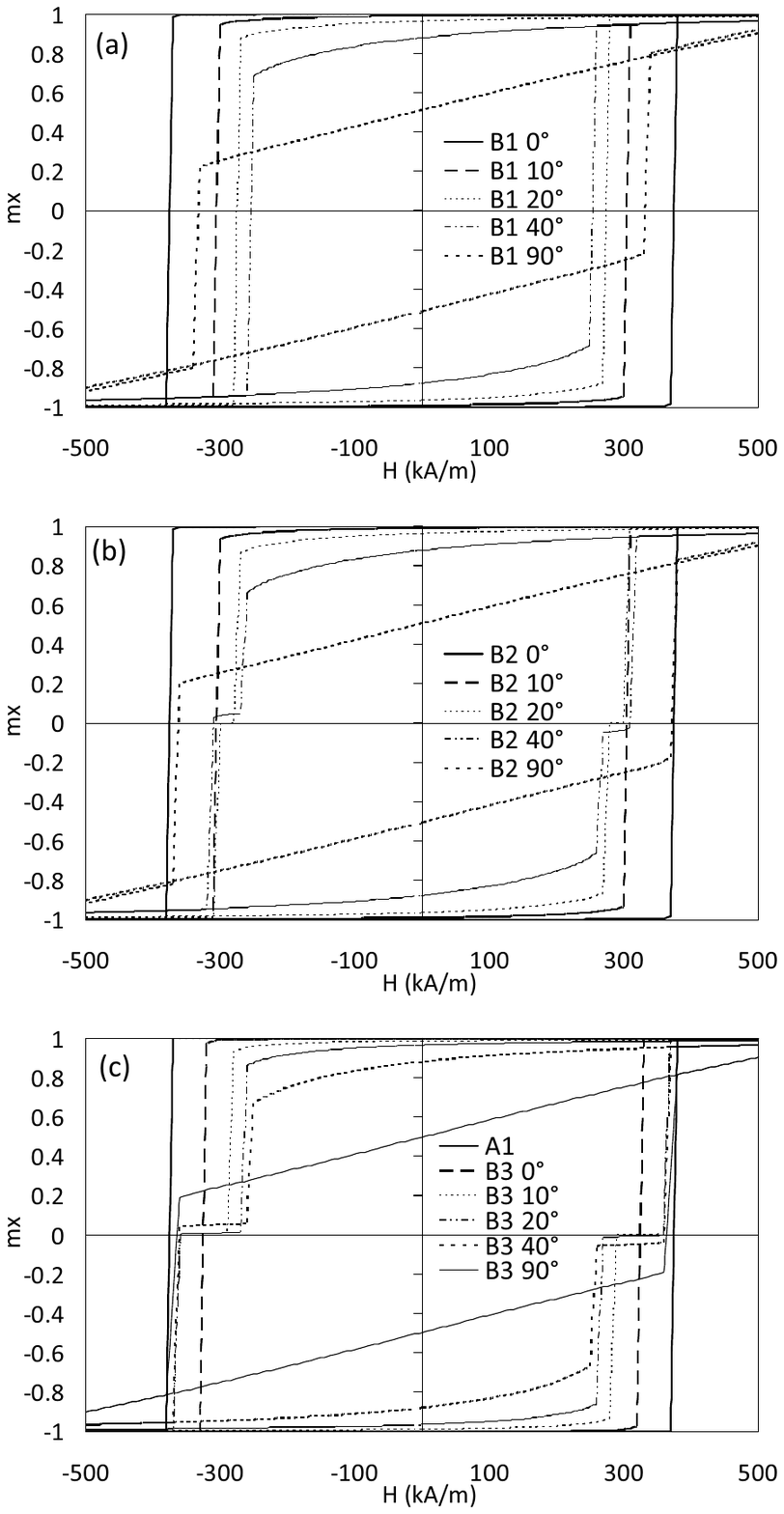}
\label{mesh} \caption{Hysteresis cycles of the (a) $B1$
configuration for $\alpha=0\text{\textdegree, 10\textdegree,
20\textdegree, 40\textdegree\ and 90\textdegree}$, (b) Hysteresis
cycles of the B2 configuration for $\alpha=0\text{\textdegree,
10\textdegree, 20\textdegree, 40\textdegree\ and 90\textdegree}$,
(c) Hysteresis cycles of the B3 configuration for
$\alpha=0\text{\textdegree, 10\textdegree, 20\textdegree,
40\textdegree\ and 90\textdegree}$.}
\end{figure}

We now investigate the effects of misorientations ($\alpha\neq0$)
for the configurations B1, B2 and B3 in order to reproduce the situation
of aggregates. In the B1 configuration, the coercive field is reduced
from $H_{C}\approx375\,\textrm{kA/m}$ for $\alpha=0\text{\textdegree}$
down to $H_{C}\approx255\,\textrm{kA/m}$ for $\alpha=40\text{\textdegree}$.
The magnetization reversal of both nanowires remains simultaneous
demonstrating that there is a strong interaction between the wires.
However in the limiting case where $\alpha=90\text{\textdegree}$
(\textit{i.e.} the nanowires are perpendicular to each other), the
coercive field is somewhat recovered to $H_{C}\approx330\,\textrm{kA/m}$
but the two wires reversals are disconnected. The magnetization of
the wire at 90\textdegree{} simply follows a linear dependance with
respect to the applied field.

The situation slightly differs in the $B2$ configurations since the
magnetization reversal of both nanowires is no more simultaneous for
$\alpha=20\text{\textdegree\ and 40\textdegree}$ showing that the
interactions between wire tips is very sensitive to geometrical details.
Furthermore, in the limiting case $\alpha=90\text{\textdegree}$,
the coercivity almost recovers the value of the isolated nanowire
case, suggesting that there are no more measurable interactions between
the wires.

The situation is qualitatively different in the $B3$ configurations.
Figure 6(c) shows that for $\alpha=10\text{\textdegree, 20\textdegree\ or 40\textdegree}$,
the magnetization reversal of both nanowires is no more simultaneous
and the coercivity of the aligned wires recovers a value of $H_{C}\approx375\,\textrm{kA/m}$
corresponding to the case of an isolated wire ($A1$). Surprisingly,
it shows that neighboring wires in the configurations $B3$ have
measurable effects on the coercive field only when they are parallel
to their neighbor and that as soon as there is a small misalignement,
there are no more interactions.

Finally, on the one hand, for perfectly aligned wires ($\alpha=0$)
, stray fields leads to a moderate loss of coercivity (13\%) only
in the $B3$ configuration. On the other hand, misalignment of the
nanowires ($\alpha\neq0$) , induces a significant loss of coercivity
in the $B1$ and $B2$ configurations while stray fields effects become
negligible in the $B3$ configurations. The dipolar interactions on
the coercivity are very sensitive to the geometrical arrangement of
wires.

\subsection{Effect of nanoscopic nucleation or pinning points}

We now consider the case of nucleation or pinning points created inside
a cigar shaped micro-structure by wires which have respectively either
a lower coercivity or a higher coercivity. In the case of a wire with
a geometrical defect, corked or with 3 endings for example, its coercivity
will be decreased. On the other hand, one may imagine that some wires
contain structural defects which pin their magnetization. For the
micro-magnetic modelling, we have thus pinned the magnetization $M_{0}$
of one of the wires of the structure in a direction along the nanowire
long axis (Fig. 7 top). Starting from a negative saturation field
$-H_{sat}$, when arriving at remanence, none of the wires has flipped
except for the pinned one. This pinned wire can thus be considered
as a nucleation point which will promote the reversal of neighboring
wires when the field is further increased. A loss of coercivity is
expected. On the other hand, starting from a positive saturation field
$+H_{sat}$, when arriving at remanence, the magnetization of all
the wires are still parallel. When the field is further decreased,
the pinned wire will not flip and will thus act as a pinning point.
A slight increase in coercivity is expected. Thus, on the calculated
hysteresis cycle, the effect of pinning can be observed on the left
side while on the right side, the effect of nucleation is observed.

\begin{figure}[h]
\includegraphics[bb=100bp 320bp 550bp 730bp,clip,width=8cm]{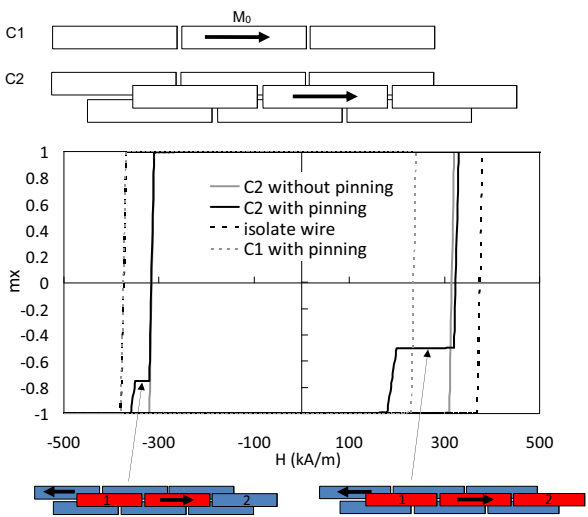}
\label{mesh} \caption{Various configurations of aggregated wires in
which the magnetization of one of the wire is pinned. (C1) Chain of
3 wires. (C2) Staggered rows of wires. }
\end{figure}

We performed micromagnetic simulations on various configurations of
aggregated wires. Let us discuss the two situations illustrated in
the geometries of Fig. 7 (top). In the first case, we look at the
behavior of a row of 3 wires in which the middle one has its magnetization
pinned. The hysteresis cycles of Fig. 7 show the evolution of the
magnetization of the free wires. That is we plot the average magnetization
(normalized to one) of all the wires except the pinned one. The reference
situation of an isolated wire is represented as the black dashed line.
When the field is swept to negative values, the coercive field of
the wires chain is unchanged with respect to the isolated wire: the
pinned wire has no effect. On the other hand, when sweeping the field
to positive values, the pinned wire acts as a strong nucleation point
and induces a 40\% drop of the coercive field with respect to an isolated
wire. The configuration C2 consists of 9 wires among which one is
pinned. The reference situation (9 free wires) is plotted on Fig.
7 as the solid grey line. The coercivity is $317\,\mathrm{kA/m}$.
In the C2 configuration, when the field is swept to negative values,
the coercive field of the wires chain is unchanged. There is only
a small step which appears in the hysteresis cycle which corresponds
to a stiffer reversal for the Nr. 1 wire (see small sketch of Figure
7). We suggest that the different behaviour of wires 1 and 2 is due
to demagnetizing field effects. When sweeping the field to positive
values, the pinned wire acts as a strong nucleation point and promotes
a first reversal of wires 1 and 2 at a field 40\% lower than the coercive
field. The other wires behavior is barely modified. The coercivity
is slightly increased which can be accounted for by dipolar interactions
with the flipped row. Thus, in conclusion, a pinned wire only affects
the behavior of its neighbor along a chain but not its side neighbors.

\section{Experimental results}

In order to probe the interactions in granular magnetic materials,
Henkel \cite{Henkel} proposed to compare the Isothermal Remanent
Magnetization (IRM) and Direct Current Demagnetization (DCD) curves.
The difference $\delta m(H)=(2.IRM(H)-DCD(H))/IRM(\infty)-1$ being
either zero in the case of non interacting particles \cite{wohlfarth1958},
positive for interactions promoting the magnetized state and negative
for interactions assisting magnetization reversal. Note that we are
measuring the DCD with an initial negative saturation field which
is opposite to the usual convention, hence the change of sign in the
previous formula for the DCD contribution. We think that this way
of measuring the DCD is more natural though.

We performed IRM and DCD measurements on aligned Co nanowires deposited
on a Si substrate (see Fig. 1). The measurements were performed both
parallel and perpendicular to the easy axis of the nanowires aggregates
(Fig. 8). One can notice that $\delta m(H)$ is mostly zero except
in 2 small regions during the reversal process. The $\delta m(H)$
value is small (max $\sim$ -14\%) and negative suggesting that only
dipolar interactions assisting reversal take place. This is in agreement
with the above calculations by showing that (i) forming aggregates
does not stabilize the magnetic structures and that only extra nucleations
can be induced by dipolar effects and (ii) that these effects remain
rather small even in very compact structures with densities close
to 1. No qualitative difference is observed in the measurements along
and perpendicular to the easy axis. The facts that two small peaks
are observed in $\delta m(H)$ during the reversal process suggests
that different mechanisms are taking place at the beginning and at
the end of the aggregate reversal. This needs to be investigated further.

\begin{figure}[h]
\includegraphics[bb=100bp 440bp 400bp 720bp,clip,width=8cm]{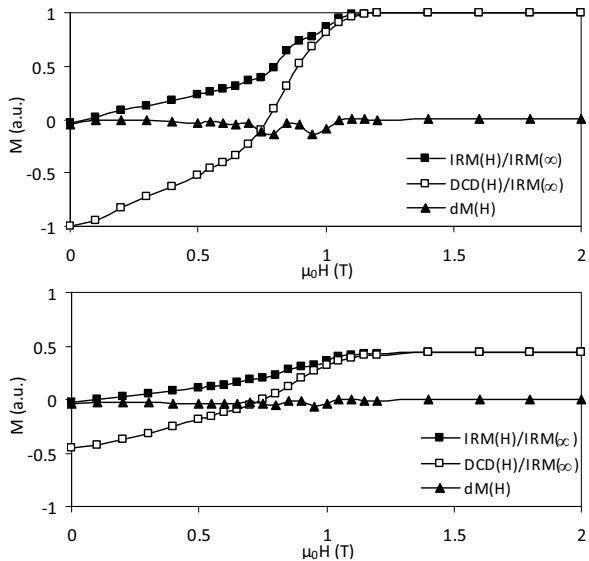}
\label{mesh} \caption{IRM, DCD and Henkel plots measured on aligned
Co wires in a polymer matrix. The magnetic field was applied either
parallel (top) or perpendicular (bottom) to the aggregates axis. }
\end{figure}

\section{Conclusion and perspectives}

We have investigated the role of magnetic dipolar interactions between
nanowires in aggregates via micromagnetic simulations. In the case
of aggregates composed of perfectly aligned nanowires, these simulations
show that no major change should be observed in the shape of the hysteresis
cycles or in the values of the coercivity and the remanence. In the
worst situation a drop of 20\% in the coercive field might be observed.
We underline that local dipolar interactions effects on the coercivity
are very sensitive to the detailled geometrical arrangement of wires.
As for the effect of nanowire misalignment inside aggregates, we show
that even very poor alignment leads only to a 30\% reduction in the
coercive field. Using Henkel plots we have experimentally confirmed
that dipolar effects are indeed small in these aggregates of nanowires
or that at least they do not significantly modify the magnetic behavior.
This validates the way of interpreting the experimental magnetic hysteresis
cycles in \cite{Maurer_APL}.

It is striking that in dense aggregates of magnetic nanowires, dipolar
interactions play a very limited role. We conclude that in dense materials based on magnetic nanowires, the
interactions between wires are negligible and that the optimization
of the materials goes through the optimization of the properties of
individual wires.
\begin{acknowledgments}
The authors gratefully acknowledge the Agence Nationale de la Recherche
for their financial support (project P-Nano MAGAFIL). We thank J.B.
Moussy (CEA-IRAMIS) for his help in the magnetometry measurements.\end{acknowledgments}


\begin{thebibliography}{23}
\bibitem{mendelsohn1955}L.I. Mendelsohn, F.E. Luborsky, T.O. Paine,
J. Appl. Phys. \textbf{26}, 1274-1280 (1955).

\bibitem{falk1966}R.B. Falk, J. Appl. Phys. \textbf{37}, 1108-1112
(1966).

\bibitem{craik1967}D.J. Craik, R. Lane, J. Appl. Phys., \textbf{18},
1269-1274 (1967).

\bibitem{AlNiCo}R. O'Handley, Modern Magnetic Materials, Principles
and Applications, p.480 (Wiley Intersciences, New York, 2000).

\bibitem{Ung2005}D. Ung, G. Viau, C. Ricolleau, F. Warmont, P. Gredin,
and F. Fi\'{e}vet, Adv. Mater., \textbf{17}, 338 (2005).

\bibitem{Ung2007}D. Ung, Y. Soumare, N. Chakroune, G. Viau, M.-J.
Vaulay, V. Richard, F. Fi\'{e}vet, Chem. Mater., \textbf{19}, 2084 (2007).

\bibitem{Soumare2008} Y. Soumare, J.-Y. Piquemal, T. Maurer, F. Ott,
G. Chaboussant, A. Falqui and G. Viau, J. Mater. Chem. \textbf{18},
5696-5702 (2008).

\bibitem{Soumare2009_Co}Y. Soumare, C. Garcia, T. Maurer, G. Chaboussant,
F. Ott, F. Fi\'{e}vet, J.-Y. Piquemal and G. Viau, Adv. Func. Mater. \textbf{19,
}1-7 (2009).

\bibitem{Soulantica09}K. Soulantica, F. Wetz, J. Maynadi\'{e}, A. Falqui,
R.P. Tan, T. Blon, B. Chaudret and M. Respaud, Appl. Phys. Lett \textbf{95},
152504 (2009).

\bibitem{Maurer_APL}T. Maurer, F. Ott, G. Chaboussant, Y. Soumare,
J.-Y. Piquemal and G. Viau, Appl. Phys. Lett. \textbf{91}, 172501
(2007).

\bibitem{EncinasPRB01}A. Encinas-Oropesa, M. Demand, L. Piraux, I.
Huynen and U. Ebels, Phys. Rev. B \textbf{63}, 104415 (2001).

\bibitem{Zighem2011}F. Zighem, T. Maurer, F. Ott and G. Chaboussant,
Journal of Applied Physics, \textbf{109}, 013910 (2011).

\bibitem{NielschAPL}K. Nielsch, R. B. Wehrspohn, J. Barthel, J. Kirschner,
U. Gosele, S. F. Fischer and H. Kronmuller., Appl. Phys. Lett. \textbf{79},
1360-1362 (2001).

\bibitem{Dobrynin_APL09}A. N. Dobrynin, V. M. T. S. Barthem and D.
Givord, Appl. Phys. Lett. \textbf{95}, 052511 (2009) .

\bibitem{Henkel}O. Henkel, Phys. Status Solidi \textbf{\emph{7,}}
919-929 (1964).

\bibitem{Garci_Henkelplot}J. Garcia-Otero, M. Porto and J. Rivas,
J. Appl. Phys., \textbf{87}, 7376-7381 (2000).

\bibitem{Ott_JAP2009}F. Ott, T. Maurer, G. Chaboussant, Y. Soumare,
J.-Y. Piquemal and G. Viau, J. Appl. Phys. \textbf{105}, 013915 (2009).

\bibitem{Maurer_PRB}T. Maurer, F. Zighem, F. Ott, G. Chaboussant,
G. Andr\'{e}, Y. Soumare, J.Y. Piquemal, G. Viau and C. Gatel, Phys. Rev.
B \textbf{80}, 064427 (2009).

\bibitem{FEMM}D. C. Meeker, Finite Element Method Magnetics. The
FEMM package is freely available at http://www.femm.info.

\bibitem{tannenwald1961} P. E. Tannenwald and R. Weber, Phys. Rev.
\textbf{121}, 715 (1961).

\bibitem{fishbacher2007}T. Fischbacher, M. Franchin, G. Bordignon
and H. Fangohr, IEEE Trans. Mag. \textbf{43}, 2896 (2007), URL : http://nmag.soton.ac.uk/nmag/

\bibitem{NetGen}NETGEN, URL: http://www.hpfem.jku.at/netgen/

\bibitem{wohlfarth1958}E.P. Wohlfarth, J. Appl. Phys. \textbf{29},
595 (1958).
\end{thebibliography}
\end{document}